 \input phyzzx\def\e{\adveq\eqno{\rm (\chapterlabel\the\equanumber)}}

\def\adveq{\global\advance\equanumber by 1}
\def\myeq{{\rm \chapterlabel\the\equanumber}}
\def\rarrow{\rightarrow}

\def\manyeq#1{\eqalign{#1}\e}

\def\semidirect{\mathrel{\raise0.04cm\hbox{${\scriptscriptstyle |\!}$
\hskip-0.175cm}\times}}


\def\ref#1{$^{[#1]}$}

\def\r#1{$[\rm#1]$}

\def\Tr{\mathop{\rm Tr}\limits}
\def\e{\adveq\eqno{\rm (\chapterlabel\the\equanumber)}}

\def\adveq{\global\advance\equanumber by 1}
\def\myeq{{\rm \chapterlabel\the\equanumber}}
\def\rarrow{\rightarrow}

\def\manyeq#1{\eqalign{#1}\e}

\def\semidirect{\mathrel{\raise0.04cm\hbox{${\scriptscriptstyle |\!}$
\hskip-0.175cm}\times}}


\def\ref#1{$^{[#1]}$}

\def\r#1{$[\rm#1]$}

\def\Tr{\mathop{\rm Tr}\limits}

\date{November, 2008}
\date{November, 2008}
\titlepage
\title{On Conformal Field Theories with Low Number of Primary Fields}
\author{Roman Dovgard and Doron Gepner}
\line{\hfill Department of Particle Physics\hfill}
\line{\hfill Weizmann Institute\hfill}
\line{\hfill Rehovot 76100, Israel\hfill}
\abstract
Using Verlinde formula and the symmetry of the modular matrix we describe an algorithm
to find all conformal field theories with low number of primary fields. We employ the
algorithm on up to eight primary fields. Four new conformal field theories are found which do
not appear to come from current algebras. This supports evidence to the fact that
rational conformal field theories are far richer than suspected before.
\endpage

Two dimensional conformal field theory has been a source of 
numerous results owing to its considerable solvability, starting
from the seminal works 
\REF\BPZ{A.A. Belavin, A.M. Polyakov and A.B. Zamolodchikov, Nucl. Phys. B 241 (1984) 333.}
\REF\Witten{E. Witten, Commun. Math. Phys. 92 (1984) 455.}
\REF\KZ{V. Knizhnik and A.B. Zamolodchikov, Nucl. Phys. B 247 (1984) 83.}
\REF\Cardy{J.L. Cardy, Nucl. Phys. B 270 [FS 16] (1986) 186.}
\REF\GepWit{D. Gepner and E. Witten, Nucl. Phys. B 278 (1986) 493.}
\REF\FQS{D. Friedan, Z. Qiu and S. Shenker, Phys. Lett. B 151 (1985) 37.}
\REF\Verlinde{E. Verlinde, Nucl. Phys. B 300 (1988) 360.}
\REF\Vafa{C. Vafa Phys. Lett. B 206 (1988) 360}
ref. \r{\BPZ,\Witten,\KZ,\Cardy,\GepWit,\FQS,\Verlinde,\Vafa}. 
In these works an axiomatic
approach to conformal field theory started to emerge with the
notion of conformal data.

Consider some extended algebra $\cal A$ which is the symmetry algebra
of the theory. 
We assume that $\cal A$ contains the Virasoro algebra ref. \r\BPZ.
The Hilbert space of the theory is then some 
representation of this algebra, $\cal H$. 
The actual symmetry algebra of the theory is two copies of the algebra
$\cal A$, ${\cal A}\times{\cal A}$, which act, respectively,
on the left and right movers.  
The Hilbert space then decomposes into 
a sum of irreducible representations 
$${\cal H}=\oplus_{i,j} N_{ij}
{\cal H}_i\times {\bar {\cal H}}_j,\e$$
where ${\cal H}_i$ denotes
the irreducible representation. 
We shall assume that the number of irreducible representations is
finite. This is called rational conformal field theory.
The $N_{ij}$ are some non--negative integers
which count how many times a representation appears in the theory.
The idea of conformal data is that these representations obey some 
relations among them which do not depend on the details of the 
symmetry algebra $\cal A$. These relations are common to a great deal
of different algebras and form some kind of universality class among
the different CFT.

We will concentrate here mainly on two main such relations, 
the modular transformations and the fusion rules along with 
the dimensions, $\Delta_i$ (mod $1$) and central charge $c$ (mod $8$).
Other relations exist which will not be discussed here, e.g., the
braiding matrix (see for example 
\REF\MS{G. Moore and N. Seiberg, RU-89-32, YCTP-P13-89, 
Trieste spring school (1989) 1.}
\r\MS\ and references within).

The partition function of the theory 
is defined by the path integral on the torus with modulus $\tau$
(ref. \r{\Cardy, \GepWit}) and it is given by
\def\trace{\mathop{\rm Trace}}
$$Z(\tau)=\trace_{\cal H} q^{L_0-c/24} \bar q^{\bar L_0-c/24},\e$$
where $c$ is the central charge and $L_0$ and $\bar L_0$ are
the left and right dimensions, and $q=e^{2\pi i \tau}$. 
Since $Z(\tau)$ depends only on the
torus, it has to be invariant under redefinitions of the torus,
called modular transformations,
$$Z(\tau+1)=Z(-1/\tau)=Z(\tau).\e$$
We define the character of the representation ${\cal H}_p$ by
$\chi_p=\Tr_{{\cal H}_p} \exp[2 \pi i \tau (L_0-c/24)]$.
The way this relation is obeyed is that the characters themselves form a representation
of the modular group,
$$\chi_p(-1/\tau)=\sum_q S_{pq} \chi_q(\tau),\e$$
where $S$ is some unitary matrix, along with the relation,
$$\chi_p(\tau+1)=e^{2\pi i(\Delta_p-c/24)} \chi_p(\tau).\e$$
Then, using the characters the partition function becomes,
$$Z(\tau)=\sum_{p,q} N_{p,q} \chi_p(\tau) \chi_q(\tau)^*.\e$$
The fact that $S$ is a unitary matrix implies that we always have the,
so called, diagonal modular invariant,
$$N_{pq}=\delta_{pq}.\e$$
The modular matrix $S$ forms part of our conformal data, where there can be many different 
theories with the same modular matrix.

Another important notion in the conformal data is the ring of fusion rules
ref. \r{\BPZ,\GepWit,\Verlinde}. Suppose $\phi_p$ is the highest weight field in the $p$
representation of the algebra. Then, we may consider what fields appear in the operator
product expansion (OPE) of two such fields, $\phi_p(w)$ and $\phi_q(z)$. The
OPE is then,
$$\phi_p(w) \phi_q(z)=\sum_{n,t} (w-z)^{n+\Delta_t-\Delta_p-\Delta_q} [\phi_t](z),\e$$
where $[\phi_t](z)$ denotes fields in the representation ${\cal H}_t$ of
the algebra. We thus define the fusion rules as the product in the fusion rules
ring as follows,
$$[p] \times [q]=\sum_t N_{pq}^t [t],\e$$
where $[t]$ corresponds to all the fields in the OPE, eq. (8), and the $N^t_{pq}$
are non--negative integral coefficients called the fusion coefficients. Since the
OPE is associative the fusion rules form a ring, called the fusion ring, where the unit
is the unit representation of the chiral algebra, which we denote by "$0$". 
The fusion ring is our second conformal data, since many theories can, in fact,
have isomorphic fusion rings.

Surprisingly perhaps, and very importantly, the two notions of modular matrix and 
fusion rules are
related by the so called Verlinde formula ref. \r\Verlinde. This formula gives 
the fusion coefficients in terms of the modular matrix,
$$N^t_{pq}=\sum_r {S_{r,p} S_{r,q} S_{r,t}^\dagger\over S_{r,0}}.\e$$
Since $S$ is unitary,
$$\sum_q S_{p,q} S_{q,r}^\dagger=\delta_{p,r},\e$$
it follows from eq. (10) that the modular matrix represents the fusion rules,
$$\psi_p \psi_q=\sum_t N^t_{pq} \psi_t,\e$$
where 
$$\psi_t^{(j)}={S_{t,j}\over S_{j,0}},\e$$
and $\psi_t^{(j)}$ is the $j$th eigenvalue of the fusion matrix.

Another simple, yet very powerful property of the modular matrix is its symmetry,
$$S_{p,q}=S_{q,p},\e$$
along with its square relation,
$$(S^2)_{p,q}=b_{p,q},\e$$
where $b$ denotes field conjugation, 
$$[\bar p]=\sum_q b_{pq} [q],\e$$
and $b$ is a permutation matrix of degree two, $b^2=1$. In a real theory
$b$ is the unit matrix. Otherwise it is the matrix that implements field 
conjugation.

Any ring which obeys the Verlinde formula with some symmetric unitary matrix $S$
we call symmetric affine variety. From the mathematical point of view, this condition
is very strong and there are very few symmetric affine varieties. In the following we will 
describe a algorithm for enumerating all the symmetric affine varieties for a small number
of primary fields. From our results, we believe that almost all such varieties do, in fact,
correspond to the conformal data of some conformal field theory.  

To check this we use an additional relation obeyed by the conformal data called 
Vafa equations (ref. \r\Vafa). This is a relation on the dimensions, $\Delta_r$ modulo an integer,
which also form a part of our conformal data. The equations are,
$$(\alpha_i\alpha_j\alpha_k\alpha_l)^{N_{ijkl}}=\prod_r \alpha_r^{N_{ijkl,r}},\e$$
where 
$$\alpha_r=\exp(2 \pi i\Delta_r),\e$$
$$N_{ijkl}=\sum_m N_{ij}^{\bar m} N_{kl}^m,\e$$
$$N_{ijkl,r}=N_{ij}^r N_{klr}+N_{jk}^r N_{ilr}+N_{ik}^r N_{jlr},\e$$
and we defined,
$$N_{ijr}=\sum_p b_{pr} N_{ij}^p.\e$$
$N_{ijr}$ is symmetric in all the indices.
We define the matrix $T$ as,
$$T_{pq}=\delta_{pq} e^{2\pi i(\Delta_p-c/24)},\e$$
which implements the modular transformation $\tau\rarrow \tau+1$, eq. (5).
Then the relation in the modular group implies
$$(ST)^3=b.\e$$
This is a very strong condition on the solution of the affine variety,
since there are many more equations than unknowns. Thus, we strongly believe
that all the affine varieties which are solutions to Vafa equations with
eq. (23) obeyed, do in fact, correspond to the conformal data of some conformal 
field theory. 

There are further two checks we can make for the consistency of the conformal data.
These are based on the Galois group of the rational conformal field theory, ref. 
\REF\Galois{D. Gepner, Int. J. Mod. Phys. B 22 (2008) 343;
D. Gepner, Phys. Lett. B 654 (2007) 113.}
\r\Galois.
We define the level of the theory as the least integer $N$ such that all the dimensions
obey
$$N\Delta_p={\rm integer}.\e$$
We then define the discriminant as
$$D=\prod_p S_{p,0}^{-2}.\e$$
Then the discriminant is an integer, which is a product of primes which divide the
level,
$$D=\prod_a  p_a^{n_a},\qquad p_a|N,\e$$
where $n_a$ are some positive integers.
For a proof see ref. \r\Galois.
Another property follows from the fact that the Galois group of a rational conformal 
field theory is always abelian. This implies that the elements of the modular matrix has
to be polynomials in some fixed root,
$$S_{ab}/S_{a,0}=P_{ab}(x),\e$$
where $x=A^{1/m}$ is some fixed root, $A$ is rational and $m$ is a positive integer 
(in all the examples it is a root of unity, $x=\exp(2\pi i/m)$),
and $P_{ab}$ are some polynomials with rational coefficients.
Both properties can be checked for the symmetric affine varieties that we find, and they are
indeed obeyed. This is an additional confirmation that these describe the data of some actual 
conformal field theories.

To conclude, we can classify all the data of rational conformal field theories (RCFT) by simply
enumerating all the symmetric affine varieties which obey the additional constraints
of Vafa equations and the Galois properties. These will give all the possible data for 
any RCFT. Unfortunately, it is very difficult to do this analytically, and we have
to resort to a computer algorithm. We use this algorithm to classify all the RCFT with up to
eight primary fields. The algorithm that we use is an improvement of the one already
used in ref. 
\REF\GK{D. Gepner and A. Kapustin, Phys. Lett. B 349 (1995) 71.}
\r\GK,
where RCFT with up to six primary fields were studied.

The idea is to generate all the fusion rings up to some limitation on the structure constant,
$$N_{ijk}\leq M,\e$$
where $M$ is some positive integer.
Then, we diagonalize the fusion ring to get the modular matrix, using Verlinde formula eq. (10).
Once we get the modular matrix, we check if it is symmetric. This gives all the symmetric affine 
varieties.

To be more specific, denote by "$0$" the unit primary field. We choose one special field different
from the unity, denoted by "$1$". We can then generate all the fusion coefficients of the
type $N_{1,i,j}$ which can be any symmetric matrix with non--negative integer coefficients,
with the only constraint that $N_{1,0}^j=\delta_{j,1}$. In practice, we have to limit the fusion 
coefficients up to some value $M$, eq. (28). Here we take $M$ to be equal to one. This limits 
our search to fusion rings where at least one field has fusion coefficients not exceeding one.

Now, once a candidate fusion ring is generated we can use 
$$\psi_1 \psi_j=\sum_k N_{1j}^k\psi_k,\e$$
where $\psi_i$ was defined in eq. (13). This means that the modular matrix is the normalized 
eigenmatrix, denoted by $Q$,
of the matrix $N_{1i}^j$. We still, however have the freedom of permuting the columns of the
eigenmatrix and the freedom of multiplying each column by $\pm 1$. Only after permuting the columns
with all possible permutations, and multiplying by all the possible signs, we can check if the 
resulting matrix is symmetric. Here, however, we improve the algorithm by using a "pre--filtration" 
of the ring. We compare the first row of $Q$ to each of the columns of $Q$, and if the vectors
are not the same (while ordered and up to signs) we can discard the ring. Only if the ring passes the 
pre--filtration we continue with generating all the permutations and signs.
This simple improvement accelerates considerably the speed of execution in a computer program,
allowing us to study all rings with up to eight primary fields.

There are two complications to be dealt with. First, if some of the eigenvalues of the $N_{1i}^j$,
denoted by $\lambda_j$ are degenerate, say $\lambda_s=\lambda_t$, the matrix of eigenvectors $Q$
is not uniquely determined, as we can have any linear combination of the $s$ and $t$ columns. We avoid
the problem by checking the first column (different from $s$ and $t$) against any row, 
while allowing the $s$ and $t$ columns to be
some unknown numbers. This provides pre--filtration for the degenerate case. If the ring passes this pre--filtration,
we check for the symmetry of the full matrix (with unknown $s$ and $t$ columns), and then
continue by generating another fusion coefficient $N_{2i}^j$, constraining associativity,
thus removing the degeneracy.

The second complication is if the ring is not real, i.e., we have some non--trivial field conjugation matrix,
$b$. This we deal with by generating all the possible values of $b$ and checking for the symmetry
of the modular matrix for each $b$.

Once a candidate fusion variety has been found, we proceed to solve Vafa equations (17) and $(ST)^3=b$
to get the dimensions and central charges. This way we classify all the RCFT with up to eight
primary fields. Most of the data can be seen to be not new, belonging to some current algebra model
which is already known. We do find however four new fusion rings which do not seem to correspond to any
current algebra. We have two new models with six primary fields and 
two with seven.
The conformal data of these models will be listed below. We believe that these correspond, in fact, to
some full fledged rational conformal field theories, yet to be explored.

If this conjecture is correct, this is a very important find. To this date all explicitly known RCFT
could be built from current algebras, i.e., WZW models with various constructions. This is to the
extent that a widely held belief by workers in the field was that these are the only existing RCFTs.
Our results show that this is not the case and that the realm of RCFT is indeed very rich.

1) Conformal field theory $A_6$ with $6$ primary fields.
The fusion rules are given by
\def\p#1{{\phi_#1}}
$$\manyeq{
\p1^2&=1+\p2+\p3\cr
\p2^2&=1+\p3+\p5\cr
\p3^2&=1+\p1+\p2+\p3+\p4+\p5\cr
\p4^2&=1+\p1+\p2+2\p3+2\p4+\p5\cr
\p5^2&=1+\p1+\p3\cr
\p1\p2&=\p1+\p4\cr
\p1\p3&=\p1+\p3+\p4\cr
\p1\p4&=\p2+\p3+\p4+\p5\cr
\p1\p5&=\p4+\p5\cr
\p2\p3&=\p2+\p3+\p4\cr
\p2\p4&=\p1+\p3+\p4+\p5\cr
\p2\p5&=\p2+\p4\cr
\p3\p4&=\p1+\p2+\p3+2\p4+\p5\cr
\p3\p5&=\p3+\p4+\p5\cr
\p4\p5&=\p1+\p2+\p3+\p4\cr
}$$
Solving Vafa equations, eq. (17), we get the dimensions of the fields which are, in the
order $\phi_0=1,\p1,\p2,\p3,\p4,\p5$ given by $0,-4/9,-1/9,1/3,-1/3,2/9$ modulo an integer.
The central charge, which solves eq. (23) is given by $8/3$ modulo $4$.The discriminant, eq. (25), is
given by $3^{12}$. Note that this is just one of the solutions to Vafa equations due to the fact that we can 
permute some of the rows of the modular matrix, leaving it symmetric, a phenomena called pseudo--conformal 
field theory. Other solutions will be just a multiplication by an integer factor $s$ of the dimensions and 
central charge, where $s$ has to be strange to the level, $N$, eq. (24), which is given here by $9$.

2) Conformal field theory $B_6$ with $6$ primary fields. 
Here the fusion rules are given by,
$$\manyeq{
\p1^2&=1+\p1+\p2+\p4\cr
\p2^2&=1+\p2+\p4+\p5\cr
\p3^2&=1+\p1+\p2+\p3+\p4+\p5\cr
\p4^2&=1+\p1+\p2+2\p3+2\p4+\p5\cr
\p5^2&=1+\p1+\p4+\p5\cr
\p1\p2&=\p1+\p3+\p4\cr
\p1\p3&=\p2+\p3+\p4+\p5\cr
\p1\p4&=\p1+\p2+\p3+\p4+\p5\cr
\p1\p5&=\p3+\p4+\p5\cr
\p2\p3&=\p1+\p3+\p4+\p5\cr
\p2\p4&=\p1+\p2+\p3+\p4+\p5\cr
\p2\p5&=\p2+\p3+\p4\cr
\p3\p4&=\p1+\p2+\p3+2\p4+\p5\cr
\p3\p5&=\p1+\p2+\p3+\p4\cr
\p4\p5&=\p1+\p2+\p3+\p4+\p5\cr
}$$
The dimensions are given by $0,2/7,-3/7,0,1/3,1/7$ modulo an integer, and the central charge is $-2$ modulo $4$.
The discriminant is $D=3^3 7^5$.

3) Conformal field theory $A_7$ with $7$ primary fields. The fusion rules are,
$$\manyeq{
\p1^2&=1\cr
\p2^2&=1+\p1+\p5\cr
\p3^2&=1+\p2+\p5+\p6\cr
\p4^2&=1+\p2+\p5+\p6\cr
\p5^2&=1+\p1+\p6\cr
\p6^2&=1+\p1+\p2\cr
\p1\p2&=\p2\cr
\p1\p3&=\p4\cr
\p1\p4&=\p3\cr
\p1\p5&=\p5\cr
\p1\p6&=\p6\cr
\p2\p3&=\p3+\p4\cr
\p2\p4&=\p3+\p4\cr
\p2\p5&=\p2+\p6\cr
\p2\p6&=\p5+\p6\cr
\p3\p4&=\p1+\p2+\p5+\p6\cr
\p3\p5&=\p3+\p4\cr
\p3\p6&=\p3+\p4\cr
\p4\p5&=\p3+\p4\cr
\p4\p6&=\p3+\p4\cr
\p5\p6&=\p2+\p5\cr
}$$
The dimensions are here $0,0,-1/7,3/8,-1/8,3/7,-2/7$ modulo an integer, and the central 
charge is $-2$ modulo $4$. The discriminant is $D=2^8 7^5$.

\par
4) The conformal field theory $B_7$ with seven primary fields. The
fusion rules are
\def\p#1{{\phi_#1}}
$$\manyeq{
\p1^2&=\p3+\p5\cr
\p2^2&=\p3+\p5\cr
\p3^2&=1+\p3+\p4\cr
\p4^2&=1+\p3+\p4\cr
\p5^2&=1\cr
\p6^2&=1+\p3+\p4+\p5\cr
\p1\p2&=1+\p4\cr
\p1\p3&=\p2+\p6\cr
\p1\p4&=\p1+\p6\cr
\p1\p5&=\p2\cr
\p1\p6&=\p3+\p4\cr
\p2\p3&=\p1+\p6\cr
\p2\p4&=\p2+\p6\cr
\p2\p5&=\p1\cr
\p2\p6&=\p3+\p4\cr
\p3\p4&=\p3+\p4+\p5\cr
\p3\p5&=\p4\cr
\p3\p6&=\p1+\p2+\p6\cr
\p4\p5&=\p3\cr
\p4\p6&=\p1+\p2+\p6\cr
\p5\p6&=\p6.\cr}$$ 
The dimensions are $0,1/32,1/32,1/4,-1/4,1/2,-11/32$ modulo an integer,
and the central
charge is $11/4$ modulo $4$. The discriminant is given by $D=2^{22}$.

The rest of the conformal field theories data, apart from the above new four 
 ones, can be seen to come
from current algebras, i.e., they are either WZW models or quotients of such models. This completes
the classification of all conformal field theories with up to eight primary fields.
However, due to large running time there may be additional degenerate cases for eight primary fields,
which we did not analyze completely.

Naturally, the important question of how to construct explicitly these four conjectured conformal field 
theories, arise. We know only the data for these RCFT's. We can however proceed along the lines of
ref.
\REF\Found{D. Gepner, "Foundations of rational quantum field theory", Caltech preprint 
CALT-68-1825, (November, 1992).}
\r\Found.
First we find the braiding matrix by solving the hexagon relation (e.g., \r\MS). Then we find a lattice
model by "Baxterising" the braiding matrix along the lines described in \r\Found. This will give
a solvable IRF model with the admissibility condition given by the appropriate fusion rules
(see ref. \r\Found\ for more detail). Calculating the fixed point RCFT of this lattice model will then
give an RCFT related to the original RCFT (in the appropriate regime of the model).
Work along these lines is currently in progress
\REF\BDG{A. Babichenko, R. Dovgard and D. Gepner, Work in progress.}\r\BDG, 
and we hope to report on it elsewhere.

We hope that this work will be the first step in exploring the vast richness of rational conformal field
theories. As was demonstrated here, current algebras are just a small subset of allowable conformal data.
It is left to future work to build explicitly these theories.   

\ack
We gratefully acknowledge discussions with A. Babichenko. We thank the Einstein center for support.
\refout
\bye